\documentclass[preprint]{aastex}

\newcommand{\delN}{$|$$\Delta$$N$(\ion{Na}{1})$|$ }
\newcommand{\nenh}{$n_{\rm{e}}/n_{\rm{H}}$}

\shortauthors{Andrews, Meyer, \& Lauroesch}

\shorttitle{Small-Scale ISM Structure}

\begin{document}

\title{Small-Scale Interstellar Na I Structure Toward M92}

\author{Sean M. Andrews\altaffilmark{1}, 
        David M. Meyer\altaffilmark{1}, and 
        J. T. Lauroesch\altaffilmark{1}}

\affil{Department of Physics and Astronomy, 
        Northwestern University, 2145 Sheridan Road, 
        Evanston, IL 60208}

\email{s-andrews2@northwestern.edu, davemeyer@northwestern.edu,
jtl@elvis.astro.nwu.edu}

\altaffiltext{1}{Visiting Astronomer, National Optical Astronomy Observatories.
NOAO is operated by AURA, Inc.\ under contract to the National Science
Foundation.}

\begin{abstract}
We have used integral field echelle spectroscopy with the DensePak fiber-optic 
array on the KPNO WIYN telescope to observe the central 27\arcsec\ $\times$ 
43\arcsec\ of the globular cluster M92 in the \ion{Na}{1} D wavelength region 
at a spatial resolution of 4\arcsec.  Two interstellar \ion{Na}{1} absorption 
components are evident in the spectra at LSR velocities of 0 km s$^{-1}$ 
(Cloud 1) and -19 km s$^{-1}$ (Cloud 2).  Substantial strength variations in 
both components are apparent down to scales limited by the fiber-to-fiber 
separations.  The derived \ion{Na}{1} column densities differ by a factor of 4 
across the Cloud 1 absorption map and by a factor of 7 across the Cloud 2 map. 
Using distance upper limits of 400 and 800 pc for Cloud 1 and Cloud 2, 
respectively, the absorption maps indicate structure in the ISM down to scales 
of 1600 and 3200 AU.  The fiber-to-fiber \ion{Na}{1} column density differences 
toward M92 are comparable to those found in a similar study of the ISM toward 
the globular cluster M15.  Overall, the structures in the interstellar 
components toward M92 have significantly lower column densities than those 
toward M15.  We interpret these low column density structures as small-scale 
turbulent variations in the gas and compare them to the larger-scale, higher 
column density variations toward M15, which may be the hallmarks of actual 
\ion{H}{1} structures.
\end{abstract}


\section{Introduction}
A significant number of studies within the last decade have served to reinforce
 the case for sub-parsec structure within the diffuse interstellar medium 
(ISM).  A longstanding method for investigating this phenomenon has been 
\ion{H}{1} 21 cm interferometric observations toward extragalactic sources 
\citep{dia89,fai98}.  Another technique uses multiepoch observations toward 
pulsars as they travel a few tens of AU each year through the ISM 
\citep{fra91,fra94}.  These observations have repeatedly shown variations down 
to scales of $\approx$ 10 - 100 AU.  In the optical, studies have focused on 
individual stars in globular clusters \citep{bat95} and binary star systems 
\citep{bla96,wat96,lau98,lau99}, typically using the \ion{Na}{1} D and 
\ion{K}{1} lines.  These observations have shown structure on scales of 
$\approx$ 10$^2$ - 10$^4$ AU for binary stars and $\approx$ 10$^4$ - 10$^6$ AU 
for globular clusters.  Additionally, several recent studies have uncovered 
evidence for temporal variations in \ion{Na}{1}, \ion{K}{1}, and \ion{Ca}{2} 
column densities toward individual and binary stars, suggesting structure at 
scales down to $\approx$ 10 - 100 AU \citep{cra00,lau00,pri00}.

These \ion{H}{1}, \ion{Na}{1}, \ion{Ca}{2}, and \ion{K}{1} observations infer 
apparently dense concentrations of atomic gas ($n_{\rm{H}}$ $\geq$ 10$^{3}$ 
cm$^{-3}$) appearing in diffuse sightlines.  The standard \citet{mck77} 
three-phase ISM theory does not sufficiently support the pervasiveness of this 
structure because the large overpressures it would exert on the ambient medium 
violates the assumption of pressure equilibrium.  In one effort to remedy the 
situation, \citet{hei97} has hypothesized a solution involving sheets and 
filaments of gas that maintain pressure equilibrium, yet could produce 
significant differences in column density measurements and the illusion of 
rather high volume densities.  \citet{elm97} offers another interpretation 
which eliminates the necessity of pressure equilibrium by suggesting that a 
fractal ISM results from turbulence, allowing self-similar structure to exist 
down to AU scales.  Other theories propose that self-gravitating, dense, 
AU-sized cloudlets may be a common neutral component of the Galactic ISM 
\citep{wal98,dra98}.

Integral field spectroscopy can be used to gain a two-dimensional perspective 
of this small-scale structure.  A recent study used this technique to map the 
ISM structure toward the globular cluster M15 \citep{mey99}.  Just as 
\ion{H}{1} observations toward extragalactic sources offer two-dimensional 
views of similar structure observed in one dimension toward pulsars, these 
integral field spectroscopic observations toward globular clusters provide 
two-dimensional maps of similar structure observed in one dimension toward 
binary stars.  The globular cluster M92 ($d$ = 8.6 kpc; $v_{\rm{LSR}}$ = -103 
km s$^{-1}$) was selected to continue this study.  Past observations of 
individual stars in M92 show interstellar \ion{Na}{1} absorption with strengths
 varying by a factor of 4 on scales down to 4\arcmin\ \citep{lan90,pil98}.  In 
this Letter, we present two-dimensional maps of the \ion{Na}{1} absorption 
toward the central 27\arcsec\ $\times$ 43\arcsec\ of M92 in an effort to probe 
the small-scale ISM structure in another Galactic halo sightline. 

\section{Observations and Data Reduction}
We used the DensePak fiber-optic array and the Bench spectrograph on the 3.5 m 
WIYN\footnote{The WIYN Observatory is a joint facility of the University of 
Wisconsin-Madison, Indiana University, Yale University, and the National 
Optical Astronomy Observatories.} telescope at Kitt Peak National Observatory 
to obtain observations of M92 in 2000 April.  DensePak is an integral field 
spectroscopy unit, consisting of 91 fibers arranged to give a sky coverage of 
27\arcsec\ $\times$ 43\arcsec\ with a fiber (3\arcsec\ diameter) 
center-to-center separation of 4\arcsec\ at the WIYN f/6.4 Nasmyth focus 
\citep{bar98}.  The DensePak array was oriented in a north-south direction 
across the core of the globular cluster, with the central fiber set at the 
center of M92 [R.A. = 17$^{\rm{h}}$17$^{\rm{m}}$07\fs3, decl. = 
+43\degr08\arcmin11\farcs5 (J2000.0)].  The configuration of the spectrograph 
was the same as described by \citet{mey99}, providing spectral coverage from 
5725 to 5975 \AA\ at a resolution of 14 km s$^{-1}$ ($R$ $\sim$ 20,000).

Three 1200 s exposures were taken of M92 with sky conditions characterized by 
$\approx$ 1\arcsec\ seeing.  The raw CCD frames were bias-corrected, 
flat-fielded, sky-subtracted (in two separate stages, the continuum and the 
terrestrial emission lines, using a 1200 s exposure of adjacent blank sky), 
combined, and wavelength-calibrated using the NOAO IRAF data reduction 
software.  Previous observations enabled us to assume that the uncertainty in 
these spectra caused by scattered-light effects should be negligible 
\citep{mey99}.  Five broken fibers and eight more with low counts were not 
used, leaving 78 of the 91 fibers providing spectra with signal-to-noise ratios
 ranging from 30 in some edge fibers to over 80 near the center.  The very weak
 telluric absorption near the \ion{Na}{1} D$_2$ $\lambda$5889.951 and D$_1$ 
$\lambda$5895.924 lines was removed by dividing the spectra by an atmospheric 
template based on observations of several rapidly rotating early-type stars 
with little intervening interstellar matter.  

Figure 1 exhibits a detailed view of the final \ion{Na}{1} spectra 
corresponding to the center of M92 and three other positions in the array.  
\begin{figure}
\plotone{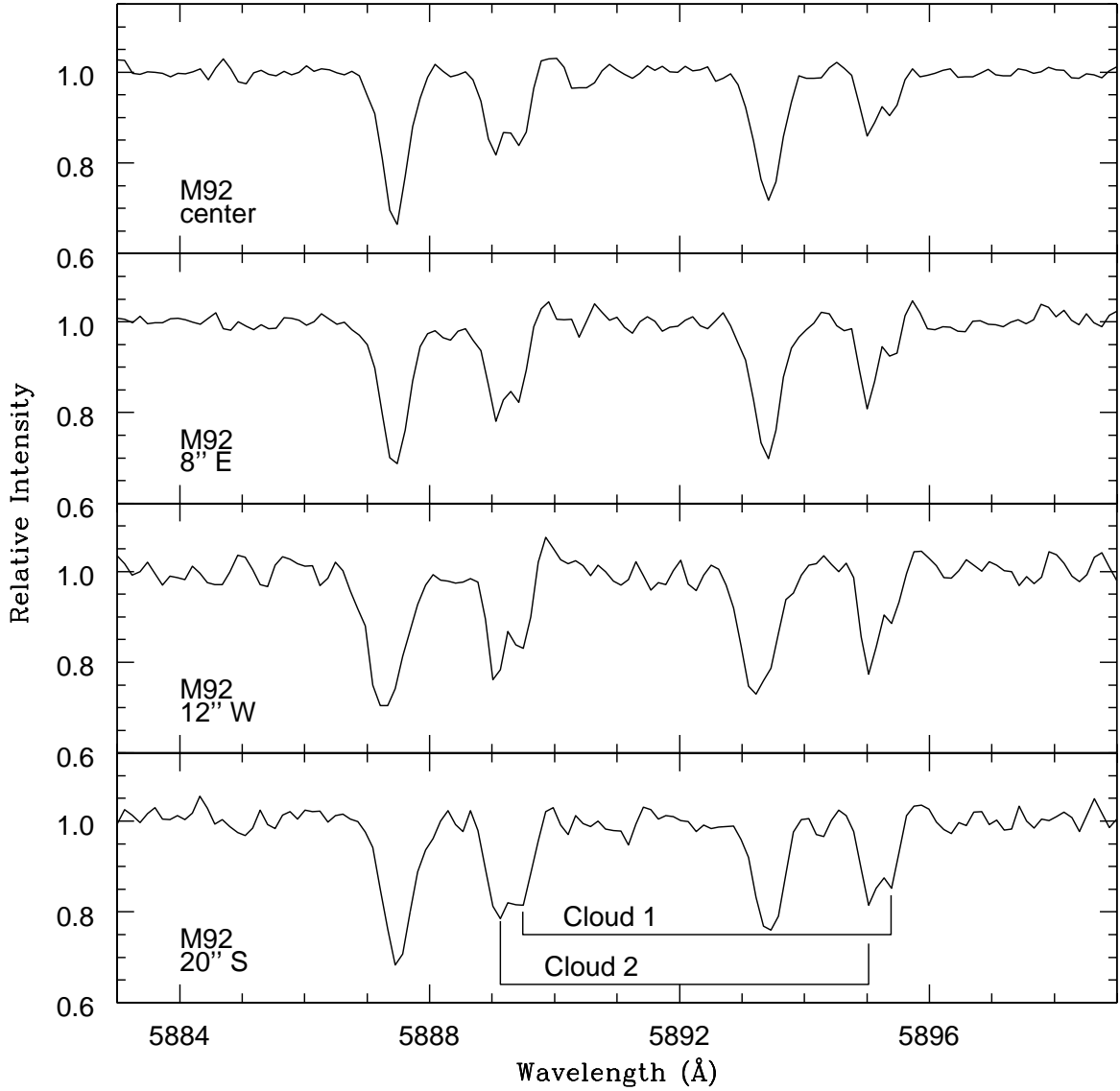}
\caption{WIYN DensePak spectra of the \ion{Na}{1} D$_2$ $\lambda$5889.951 and 
D$_1$ $\lambda$5895.924 region toward the core of M92 and three labeled 
positions of various angles and separations from the center.  The spectra have 
a velocity resolution of 14 km s$^{-1}$ and are displayed on a geocentric 
wavelength scale.  Three \ion{Na}{1} doublets appear, with the interstellar 
lines strongly blended: the bluemost is due to stellar \ion{Na}{1} absorption 
in M92, the middle component (Cloud 2) is due to interstellar gas at 
v$_{\rm{LSR}}$ = -19 km s$^{-1}$, and the redmost component (Cloud 1) is due to
 gas at v$_{\rm{LSR}}$ = 0 km s$^{-1}$.}
\end{figure}
Three \ion{Na}{1} doublets appear, with the interstellar lines strongly 
blended: the bluemost is due to stellar \ion{Na}{1} absorption in M92, the 
middle component (Cloud 2) is due to interstellar gas at v$_{\rm{LSR}}$ = -19 
km s$^{-1}$, and the redmost component (Cloud 1), is due to gas at 
v$_{\rm{LSR}}$ = 0 km s$^{-1}$.  Figure 1 also shows evidence that both 
absorption components vary on scales much less than 1\arcmin.  Over the entire 
DensePak array, the \ion{Na}{1} D$_1$ line equivalent widths vary from 30 to 
100 m\AA\ for Cloud 1 and 15 to 115 m\AA\ for Cloud 2.  Column density 
measurements were determined using the profile-fitting software XVOIGT 
\citep{mar95} by fitting the D$_2$ and D$_1$ lines simultaneously in each 
fiber, assuming single-component Voigt profiles for both of the clouds' 
\ion{Na}{1} doublets.  The line widths ($b$-values) were typically around 2 km 
s$^{-1}$ and 1.5 km s$^{-1}$ for Cloud 1 and Cloud 2, respectively.  The column
 density fits have formal profile-fitting uncertainties of about 10\% - 20\%.   

The assumption that Cloud 1 and Cloud 2 each consist of a single ISM component 
is a simplification.  Although we cannot rule out multiple component 
structure, the overall effect on the small-scale structure analysis should be 
insignificant because the \ion{Na}{1} lines are weak enough so that the 
equivalent width measurement errors should dominate the uncertainties.  
Furthermore, there are no systematic velocity variations in Cloud 1 and Cloud 2
 across the DensePak array, which reinforces the simplifying assumption of 
single ISM components.  Based on the F2 composite spectral type and extremely 
low metallicity ([Fe/H] = -2.29) of M92, the only possible stellar absorption 
(besides the stellar \ion{Na}{1} lines) in the vicinity of the interstellar 
\ion{Na}{1} features are the \ion{Ni}{1} $\lambda$5892.883 and \ion{Ti}{1} 
$\lambda$5899.304 lines \citep{mon99}.  Since these lines would lie about 1 
\AA\ from the Cloud 1 and Cloud 2 \ion{Na}{1} features and are not apparent in 
the spectra in any case, any stellar contamination of our measured \ion{Na}{1} 
column densities should be minimal.

\section{Discussion}
\citet{lil91} have measured interstellar \ion{Na}{1} absorption at 
v$_{\rm{LSR}}$ = 0 km s$^{-1}$ toward HD 155639 ($d$ $\approx$ 200 pc; 
2$\fdg$75 separation from M92) and somewhat stronger absorption toward HD 
157908 ($d$ $\approx$ 400 pc; 3$\fdg$75 separation from M92).  From these 
observations, we can place an upper limit of 400 pc on the distance to Cloud 
1.  In the case of Cloud 2, we first note that \citet{mun62} observed 
interstellar \ion{Ca}{2} at LSR velocities of 0 km s$^{-1}$ and -19 km 
s$^{-1}$ toward HD 156110 ($d$ $\approx$ 800 pc; 2$\fdg$4 separation from 
M92).  For confirmation, we used STIS UV echelle spectra of this star from the 
HST archive.  The \ion{C}{1} and \ion{C}{1}$^{\star}$ lines in these spectra, 
as well as the \ion{H}{1} 21 cm emission lines from the Dwingeloo Survey 
\citep{har97}, exhibit interstellar components with similar velocities.  We 
also obtained \ion{Ca}{2} and \ion{Na}{1} data from the Kitt Peak coud\'{e} 
feed telescope, which again showed the same interstellar velocity components 
toward HD 156110.  From this data we can place a distance upper limit of 800 pc
 on this cloud.

Using these distance estimates, the 27\arcsec\ $\times$ 43\arcsec\ DensePak 
array coverage corresponds to a 10,800 $\times$ 17,200 AU (0.05 $\times$ 0.08 
pc) section of Cloud 1 and a 21,600 $\times$ 34,400 AU (0.10 $\times$ 0.17 pc) 
section of Cloud 2.  The 4\arcsec\ fiber-to-fiber spacing then translates to 
1600 and 3200 AU in Cloud 1 and Cloud 2, respectively.  The derived Cloud 1 
\ion{Na}{1} column densities extend from 3.3 $\times$ 10$^{11}$ to 1.2
$\times$ 
10$^{12}$ cm $^{-2}$.  The maximum $N$(\ion{Na}{1}) variation observed across 
the 1600 AU scale is 6.5 $\times$ 10$^{11}$ cm$^{-2}$, and the median \delN is 
1.2 $\times$ 10$^{11}$ cm$^{-2}$.  Cloud 2 generally has higher \ion{Na}{1} 
column densities, with individual fiber values ranging from 4.9 $\times$ 
10$^{11}$ to 3.6 $\times$ 10$^{12}$ cm$^{-2}$.  Over the 3200 AU scale, the 
maximum $N$(\ion{Na}{1}) variation in this cloud is 2.9 $\times$ 10$^{12}$ 
cm$^{-2}$, and the median \delN is 3.3 $\times$ 10$^{11}$ cm$^{-2}$.

The M92 data presented here introduces some interesting differences and 
similarities with M15, the only other globular cluster studied in this way to 
date \citep{mey99}.  First, the observations were performed slightly 
differently, with the fainter M92 receiving only 70\%\ of the exposure time of 
M15 due to poor weather, resulting in lower signal-to-noise ratios for M92.  
The interstellar components toward M92 are only somewhat resolved, whereas the 
M15 clouds are well separated in velocity.  Aside from these issues, the final 
maps are directly comparable.  The M92 and M15 absorption maps presented in 
Figure 2 show a striking difference, but also an important similarity.  
\begin{figure}
\plotone{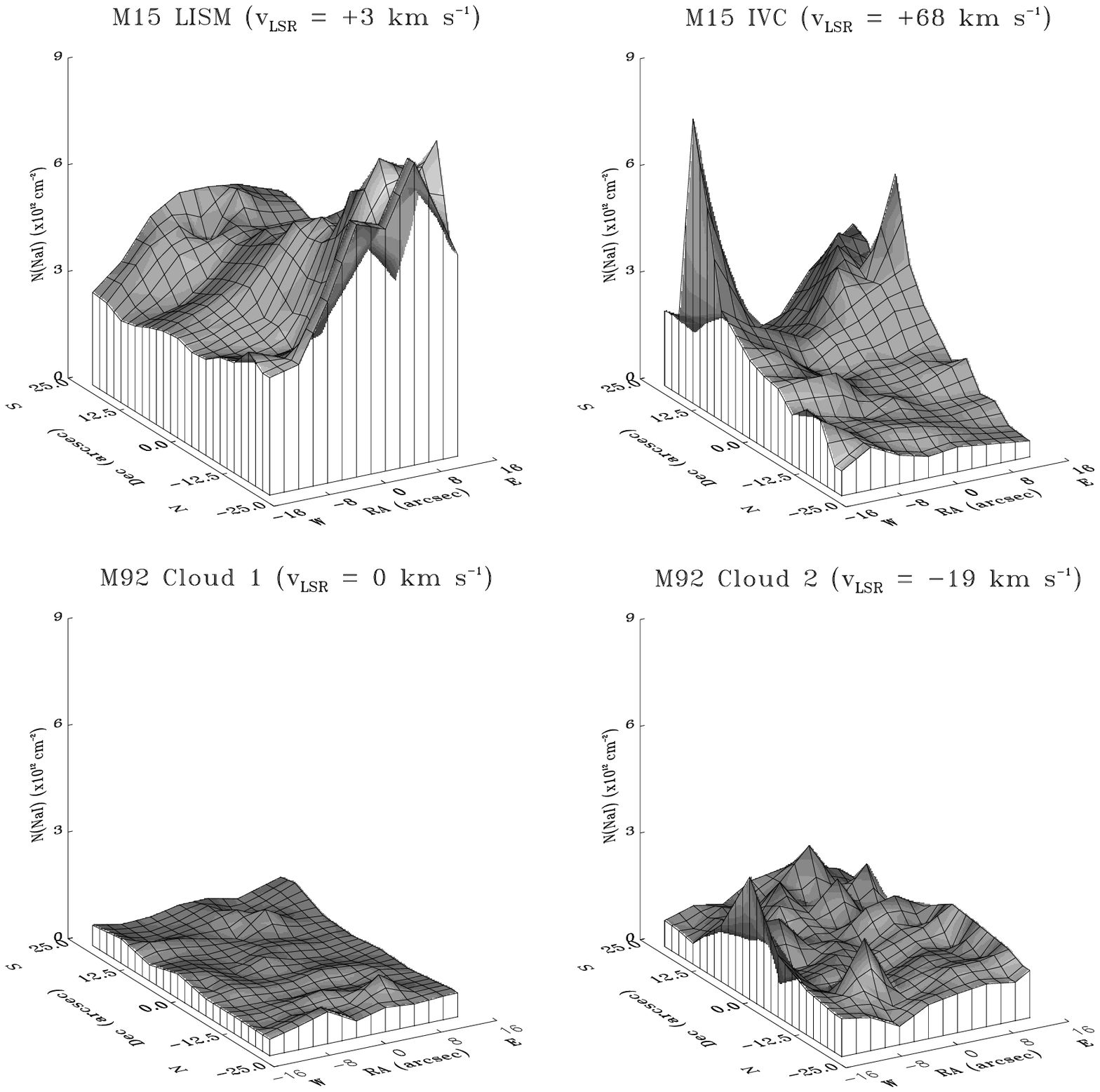}
\caption{Surface maps of the \ion{Na}{1} column densities 
corresponding to (clockwise from the upper left) the v$_{\rm{LSR}}$ = +3 km 
s$^{-1}$ LISM component toward M15, the v$_{\rm{LSR}}$ = +68 km s$^{-1}$ IVC 
component toward M15 (the M15 maps were reproduced from \citet{mey99}), the 
v$_{\rm{LSR}}$ = 0 km s$^{-1}$ Cloud 1 component toward M92, and the 
v$_{\rm{LSR}}$ = -19 km s$^{-1}$ Cloud 2 component toward M92.  The \ion{Na}{1}
 column densities were derived from individual fiber spectra such as those 
displayed in Figure 1 using a single-component Voigt profile fit.  Since the 
centers of alternating rows are offset by a half-fiber in a honeycomb 
configuration, the column densities were put into a 14 $\times$ 13 array by 
interpolating between points in RA.  This surface plot was generated from a 
rebinning of this array and has a spatial resolution of about 4\arcsec.}
\end{figure}
Unlike 
the M92 maps, the M15 maps show large, high column density features.  However, 
the low column density regions of the M15 maps are virtually identical to the 
M92 maps.  Regardless of the contrasts between the maps, the variations toward 
M92 are appreciably larger than the column density measurement errors, thus 
ruling out the possibility that they are merely random noise fluctuations.  For
 example, the NW feature in the Cloud 2 component toward M92 (lower right map 
of Figure 2) has a column density enhancement relative to the local \ion{Na}{1}
 baseline that is approximately 4.5 times the measurement uncertainty.  The 
similarities between the maps suggest that the M92 variations are typical in 
low \ion{Na}{1} column density components.  Even though the M15 columns are 
generally higher than those of M92, and have a greater dynamic range in 
$N$(\ion{Na}{1}), the median fiber-to-fiber column density differences are 
comparable, statistically reaffirming that we have measured similar variations 
in lower column density gas.  

The physical interpretation of these variations is uncertain.  The 
possibilities are that the observed \ion{Na}{1} structures are caused by real 
\ion{H}{1} structures or fluctuations in the physical conditions of the gas.  
The former interpretation is speculative as \ion{Na}{1} is not a dominant ion 
in \ion{H}{1} clouds, and therefore does not necessarily trace \ion{H}{1} 
structure.  While it has been shown that empirical estimates of $N$(\ion{H}{1})
 from $N$(\ion{Na}{1}) can be made when diffuse Galactic clouds have 
$N$(\ion{Na}{1}) $\gtrsim$ 10$^{12}$ cm$^{-2}$ \citep{hob74,wel94}, the use of 
this relationship in the case of any structures in the M92 maps would be 
somewhat dubious because the column densities measured are typically below or 
comparable to this limit.  The latter interpretation relies on fluctuations in 
the Na ionization equilibrium, or equivalently the $N$(\ion{Na}{1})/$N$(H) 
ratio.  \citet{lau98} discovered that $N$(\ion{Na}{1}) and other neutral 
species vary toward the binary $\mu$ Cru, whereas \ion{Zn}{2}, a dominant ion, 
does not.  This suggests that interstellar $N$(\ion{Na}{1}) variations could be
 the result of small-scale perturbations in the Na ionization equilibrium, 
caused by density and/or \nenh\ fluctuations.

The comparison between the M92 and M15 maps provides a new perspective on the 
\ion{Na}{1} structure.  Namely, the variations toward M92 and the low column 
density regions of M15 may be small-scale undulations in the gas structure due 
to turbulence or other variations in physical conditions, whereas the larger 
features in the high column density regions of the M15 maps may be 
representative of actual \ion{H}{1} structures.  Utilizing integral field 
spectroscopy to continue to map ISM structure in both low and high \ion{Na}{1} 
column density components toward other extended background sources will
help to 
test this idea.  

To summarize, our observations show that two components of the ISM toward M92 
exhibit significant structure down to scales of a few arc seconds, which could 
be indicative of variations in \ion{H}{1} column densities, and/or turbulent 
fluctuations in the physical conditions of low column density interstellar 
components.  The complimentary low column density features among the M92 and 
M15 maps uncovers an interesting possible interpretation of these structures 
and the relationships between the larger variations toward M15 and \ion{H}{1}
 structures.  With continued absorption-line mapping of interstellar components
 toward these and other globular clusters using instruments like DensePak, it 
is possible to further examine the spatial structure of diffuse Galactic clouds
 over a wide range of physical scales.  Additionally, as recently shown by 
\citet{rot00}, this same technique can probe similar structure using galaxies 
as background sources, perhaps providing interesting new information about 
high-velocity clouds or the ISM in other galaxies.

\acknowledgments
We are grateful to the referee for his/her valuable comments, and also to Dan 
Welty, Stefan Cartledge, and Matt Haffner for providing \ion{Ca}{2}, 
\ion{Na}{1}, and \ion{H}{1} spectra of HD 156110, respectively.  S. M. A. 
would also like to thank Dan Welty and Bruce Draine for useful conversations, 
and Caty Pilachowski for providing additional \ion{Na}{1} spectra of M92.

\end{document}